# Influence of Electromagnetic Fields on Nuclear Processes

A. Y. Wong*, J. Z. Chen, M. J. Guffey, A. Gunn, B. Mei, C.C. Shih, Q. Wang and Y. Zhang

Alpha Ring International, Ltd., Monterey, California 93940

*Correspondence author: awong@alpharing.com

11/4/2020

**Abstract**

Although the energies associated with nuclear reactions are due primarily to interactions involving nuclear forces, the rates and probabilities associated with those reactions are effectively governed by electromagnetic forces. Charges in the local environment can modulate the Coulomb barrier, and thereby change the rates of nuclear processes. Experiments are presented in which low-temperature electrons are attached to high-density rotating neutrals to form negative ions.

The steady-state quiescent rotating plasma generated here lends itself to first prove the principle that low temperature systems can yield MeV fusion particles. It allows the use of high density of neutrals interacting with the wall to yield gain greater than unity. It also demonstrates that instabilities can be avoided with high neutral densities. Collective dynamic interactions within this steady-state quiescent plasma result in an arrangement of negative charges that lowers the effective Coulomb barrier to nuclear reactions at a solid wall of reactants. MeV alpha particles are synchronously observed with externally imposed pulses as evidence of fusion being enabled by Coulomb fields. Impacts on fusion, the source of energy in the universe, will be discussed.

# INTRODUCTION

Nuclear forces create large potential energy states over a short distance range relative to the much weaker electromagnetic (EM) forces that act over long distances. The nuclear potential energy alone essentially determines the overall thermodynamics and stability of nuclear systems; however, the rates of fusion and alpha decay are governed primarily by EM forces. In particular, the rates of these thermodynamically favored processes are limited by the potential energy surface known as the Coulomb barrier. An enhancement of the fusion rate by proximal negative charges has been demonstrated as muon catalyzed fusion observed by Alvarez [1]. The muon, as a single negative charge of heavy mass, can increase the effective fusion cross section, but a system dependent on such a short-lived particle (2 µs) has not been found to be a practical energy source. In our method we are substituting the field of a muon by a large number of long-lived negative hydrogen ions in the vicinity of many fusion pairs.

Conventional nuclear fusion efforts are generally concentrated at overcoming the Coulomb barrier through the confinement of energetic ions of tens to hundreds of keV. We propose that the Coulomb barrier, created by repulsion between positive charges, can however, be reduced by suitably arranged negative charges. Collective oscillating fields due to the coherent motion of multiple electrons, either free or residing in multiple negative hydrogen ions, can enhance the fusion probability by lowering the Coulomb barrier and have a lifetime much longer than a muon. We have named this process of lowering the Coulomb barrier as "Electron Catalyzed Fusion" (ECF).

Negative hydrogen ions ($H^-$) form in an electron-rich environment by the attachment of an electron to a hydrogen atom. A neutral hydrogen atom has a natural electron affinity to acquire a free electron, releasing 0.754 eV energy in an exothermic reaction. In the presence of electron emission sources and high neutral densities, the density of negative ions H⁻ can be made comparable to the neutral density. It can be higher than the electron density according to the Saha equilibrium calculation. Our past experiments on negative ions [2,3] have shown that a plasma consisting mainly of positive and negative ions is much more stable than a plasma containing only a single ion species and electrons; the confinement time is greater by the square root of the mass ratio $M_i / m_e$.

Calculations of nuclear systems assuming a constant environment show that densities of charges large enough to have a significant effect on the Coulomb barrier are unlikely because of space-charge repulsive forces. Considering, however, a profusion of electrons and negative hydrogen ions greater than $10^{17}/cm^3$ oscillate against positively charged reactants, maintaining charge neutrality on the average, and yet producing sufficient electric fields to lower the Coulomb barrier [4] for the fast fusion process ($10^{-14}$ s) to occur. These resonant oscillations are in the 200 GHz regime and are characterized by $\omega\tau \gg 1$ and $k\lambda \gg 1$ on account of their mm wavelengths.

We have chosen the hydrogen-boron fusion reaction in equation 1 to demonstrate ECF. This aneutronic reaction has the advantage that the charged, non-neutronic fusion products generate electrons and ions which provide positive feedback to the fusion process. Theoretical order-of-magnitude estimates based on resonant electric fields generated by the oscillations between hydrogen anions formed in the plasma and boron nuclei in solid Lanthanum hexaboride (LaB$_6$), with densities estimated at ~$10^{22}/m^3$ for H⁻ and B⁺ predict that such an enhancement of fusion under moderate conditions is possible [4].

$${}^{1}_{1}H + {}^{11}_{5}B \rightarrow 3\alpha + 8.7 \; MeV \tag{1}$$

In the fusion process between oppositely charged reactants H⁻ and B⁺, fundamental plasma wave experiments have shown that large, localized potentials (e$\Phi$/kT≫1) can be generated from oscillatory motion in space and time [4–6]. This potential can partially negate the Coulomb repulsion, making quantum tunneling easier between low energy reactants. Additionally, by increasing the pressure of neutral fusion reactants, which do not cause space-charge related instabilities, the necessary densities are more easily achieved. A method of controlling the motion of neutrals through a minority group of ions was developed from the strong coupling between high-density neutrals and ions through collisions [7,8].

We report experimental observation of fusion products of MeV alphas from plasmas with temperatures in the sub-eV range. This can only happen if the Coulomb barrier is lowered, which we predict is caused by the collective effect of low energy electrons and negative hydrogen ions. Without lowering the Coulomb barrier, the threshold proton-boron center-mass energy required is ~70 keV.

## EXPERIMENT

We have developed a rotating, cylindrical low temperature plasma at high densities of hydrogen neutrals and electrons. Each apparatus consists of an inner cylindrical electrode, and an outer electrode, a cylindrical tube with open ends. An axial magnetic field, typically $B \approx 0.09$ T, is generated either by a row of neodymium-iron-boron (NdFeB) permanent annular magnets, or superconducting solenoid magnets. A voltage is applied between the inner and outer electrodes, causing a radial current $I$ to travel along the radial path distance **L** between inner (positively biased) and outer (grounded) electrodes. Neutral atoms, electrons, and positive and negative ions are formed during the initial discharge. This current consists of positive and negative hydrogen ions flowing in opposite radial directions; ions have a greater mass and mobility than electrons to cross the axial magnetic field by a factor of $(M_i/m_e)^{1/2}$. A rotation of this weakly ionized plasma is achieved by $I \mathbf{L} \times \mathbf{B}$ forces, where **L** is ~1 cm; no mechanical rotors were used. With hydrogen pressures in the torr range ($n_o \sim 10^{23}$ m$^{-3}$), where the mean free path is on the order of microns, ions and neutrals are driven to rotate together [7]. This rotating plasma is characteristic of the Alpha Ring (AR) system.

Six pieces of solid LaB$_6$ are mounted on the inner surface of the outer electrode. This thermionic material serves as both the source of electrons and the boron nuclei reactant ($^{11}$B). The LaB$_6$ emit electrons strongly in the visible and infrared wavelengths consistent with being heated above ~1900 K. As a result of the centrifugal acceleration (~$2 \times 10^4$ m/s$^2$), hydrogen neutrals, and ions are moved to the outer edge of the chamber where fusion reactions can occur. We propose that these electrons infuse with the rotating plasma and shift the equilibrium to favor the production of negative hydrogen ions. This electron emission may be further enhanced when the LaB$_6$ is bombarded by MeV particles produced during fusion events.

The total pressure is maintained at a constant value between 1 to 3 Torr, monitored by a convection-enhanced Pirani gauge. Temperatures in the chamber are measured using K-type thermocouples and the relative changes in temperature are monitored by infrared and visible light emission. The temperature of the plasma is determined from measurements of the intensity ratio of the hydrogen atomic spectral lines at 656 nm and 486 nm [9]. These plasma temperatures are used to calculate the relative equilibrium ion populations using Maxwell-Boltzmann methods.

The experimental design with confinement of the hydrogen plasma at the outer electrode presents the H- with multiple opportunities to collide with $^{11}$B nuclei at or near the wall and offers easier access to position devices for measurement of fusion products. The rotation rate is generally greater than $10^3$ rotations per second and the electron densities are $n_e > 10^{21}$ m$^{-3}$ as measured by Stark broadening of the hydrogen Balmer series beta line at 486.1 nm [10]. The density of negative hydrogen ions $> 10^{23}$ m$^{-3}$ is estimated from the neutral hydrogen density of 1 Torr and the following simplified Saha equation,

$$\frac{n_-}{n_0 n_e} = \Lambda^3 \exp\left(\frac{E}{kT}\right), \tag{2}$$

where $\Lambda^3$ is the cube electron thermal de Broglie wavelength = $2\,(h^2/2\pi\,m\,k\,T)^{3/2}$, E is the energy of 0.754 eV released when the H⁻ is formed, and $T$ is the electron temperature in the rotating frame of the neutral hydrogen atom; $T$ can be 5% of the initial temperature of the electron source if all electrons rotate at the speed of the neutral atoms.

Plasma generation is initiated in an AR system when power is injected, usually at about 350 V. The power supply used in our experiments operates in a current limiting mode. Operation times are extended with active water-cooling in the inner and outer electrodes. During longer operation times, a characteristic drop in voltage followed by a steady-state, quiescent operation mode has been observed. This is indicated by a minimally oscillating voltage and conductance with noise fluctuations of ~3%, as shown Fig. 1. The minimum breakdown potential for hydrogen gas is 273 V, suggesting that in this quiescent state the dominant charge carriers are negative ions, and electrons from LaB$_6$. Once a steady-state operation is reached, the current can be pulsed at different values for correlation with MeV particles.

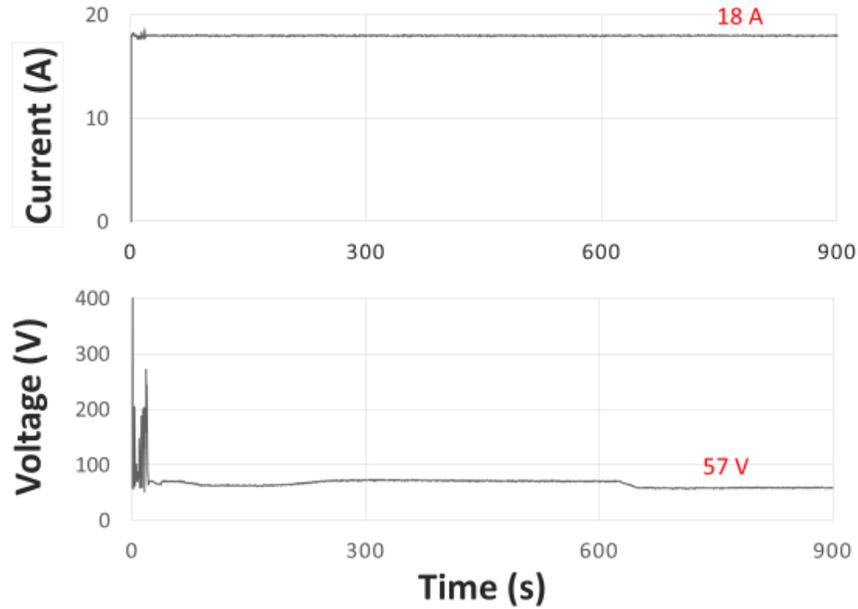

FIG. 1. A mode of steady-state quiescent operation in an AR system. The stabilization of the system voltage and current occur at ~50 seconds after initiation (Run 20190417). Operation times of hours have been performed.

The energy of all reactants measured by Langmuir probes is in the range of less than 1 eV. We believe that the energy is kept low enough to form hydrogen ions as a result of frequent collisions between ions and neutrals. Notwithstanding these low-energy conditions, we observe fusion products at MeV energies confirming that fusion has occurred. We attribute this increased fusion probability to the Coulomb barrier being lowered. During repetitive pulse operation, we detect the production of alpha particles (doubly-charged helium nuclei) as evidence of fusion. Our experimental conditions are less than 1 eV ($kT \approx 0.2$–$0.6$ eV), with reactant densities greater than $10^{23}$ m$^{-3}$ for hydrogen in the plasma, and $6.7 \times 10^{28}$ m$^{-3}$ for $^{11}$B from solid LaB$_6$ based on the average natural abundance of the isotope.

The first diagnostic of MeV fusion products is CR-39 plastic detectors [11] made from a hydrocarbon compound (Columbia Resin 39 or poly allyl diglycol carbonate). This commonly used nuclear radiation track diagnostic relies on damage to the chemical bonds by MeV particle during its penetration through the detector medium. Exposure to a 6 M NaOH solution for 5.5 hour etches the damaged portion of the detector more quickly than the undamaged portion, resulting in tracks of various depths and diameters.

Track depths for the CR-39 samples are measured using a Bruker NT9300 optical profilometer operating in vertical scanning interferometry mode with a 20× objective lens. The data obtained from this system, which is analyzed using the open-source Gwyddion software package [12] includes surface (profilometry) data as well as image data obtained near the surface focus. The image data is used to define the (2D) surface area of the track opening as the diameter, while the depth is determined from the lowest point of the defined track area in the surface profilogram, with the height at the surface of the coupon defined as zero. This profilometer was set up at the UCLA campus where the California Nanotechnology System Institute (CNSI) can provide third-party validation of data. The diameters and depths of penetration of the resulting tracks are measurable and can be compared with those simulated using the computer modeling software called TRACK_TEST [13,14] or its successor, TRACK_VISION [15]. Doubly-charged helium alphas are considered in the simulation. This detection method was calibrated with a radioactive source whose emitted alpha energies measured at the detector are controlled by the thickness of aluminum sheets in front of the source [4].

Geiger Muller (GM) detectors [16] are used to detect and measure ionizing radiation. It detects ionizing radiation such as alpha particles, beta particles, and gamma rays using the ionization effect produced in the detector. Pulses produced by the GM detectors are amplified and then counted using a LabJack T4 DAQ device. The final readout is displayed as counts per second, or number of ionizing events.

**RESULTS**

In our experiments, a power input with constant current and 500 V maximum is used, and the measured energy of all reactants are in the range of less than 1 eV, as a result of frequent collisions between ions and neutrals. Notwithstanding these low-energy conditions, we observe fusion products at MeV energies confirming that Coulomb barriers are lowered. We have chosen detection methods that are selective to energetic particles in the MeV range. These detection systems include CR-39 and Geiger-Muller detectors.

In Fig. 2, a CR-39 detector track for an experimental sample exposed to the AR plasma in a superconducting magnet chamber is compared to a simulation track in TRACK_TEST [4]. The CR-39 sample was carefully cut to reveal the depth of penetration by alphas. This commonly used nuclear radiation track diagnostic method for energetic particles does not require elaborate electronics.

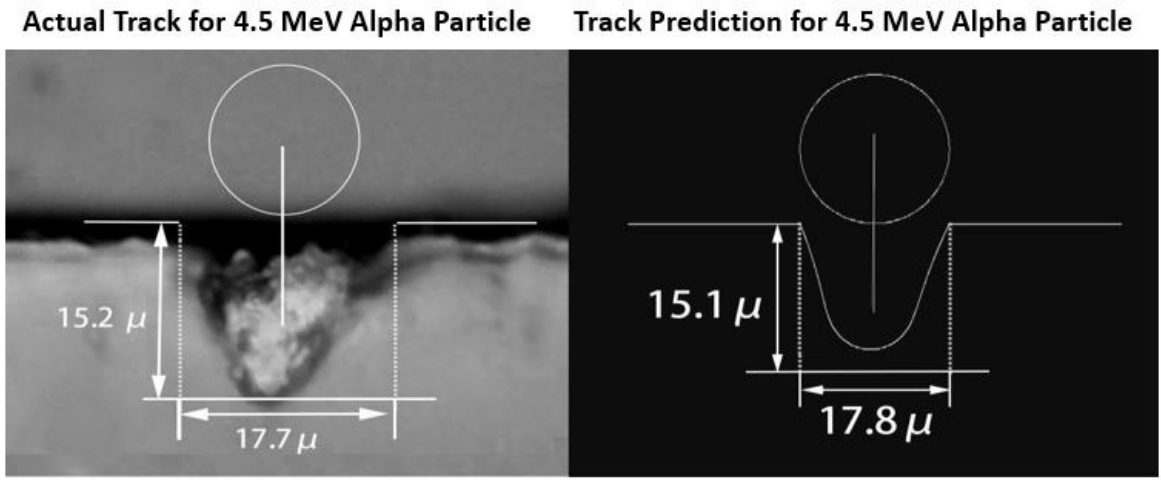

FIG. 2. Comparison between an actual alpha track from an AR plasma fusion product observed under microscope and the simulated alpha track profile as predicted by the TRACK_TEST simulation program [4,13,14].

Our measurements of various diameters and depths of tracks in CR-39 samples exposed to the rotating plasma yield an estimate of 1 to 5 MeV energies for particles produced from fusion.

The correlation of measured diameter and depth from experiments with a simulation track in TRACK_VISION [15], shown in Fig. 3, confirm that the tracks are consistent with those predicted for alpha particles with energies up to 6 MeV.

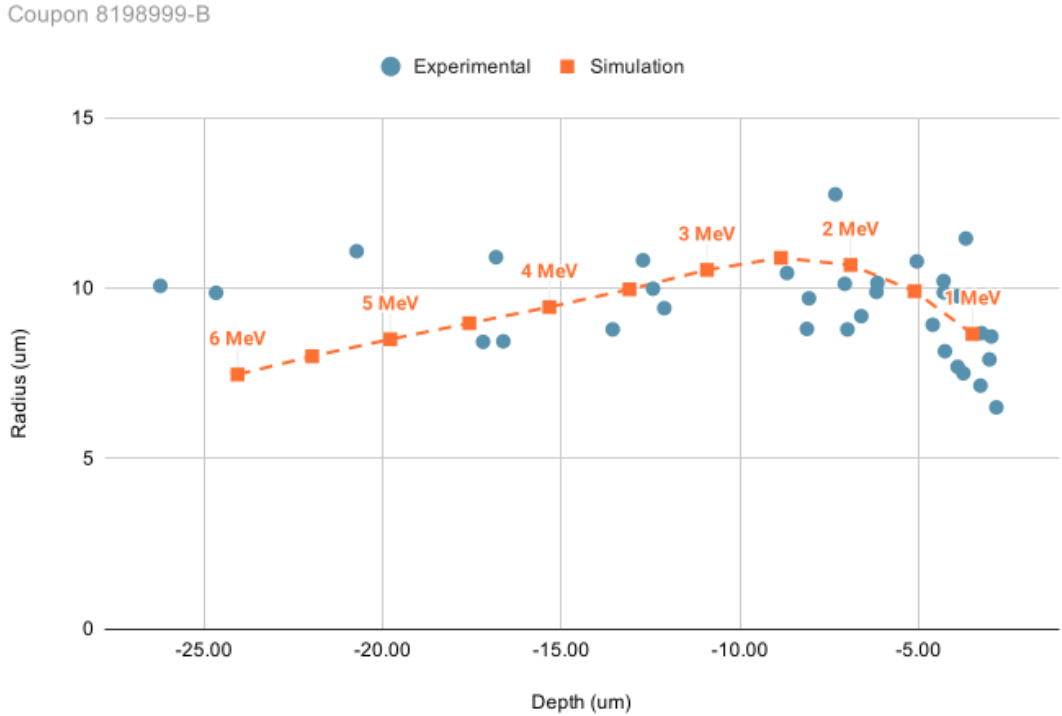

FIG. 3. Comparison between experimental and simulated track parameters. Uncertainties in experimental track radii are ± 0.9 µm. Simulations in TRACK_VISION [15] were conducted with etch rate 3 µm/hr, etch time 5.5 hr. The experimental tracks were exposed to plasma for 8 pulses of 10 s each. The current varied between 17 and 21 A shot-to-shot, and the voltage was in a range of ~200 to 500 V (Run 20200803).

The CR-39 detectors are essentially immune to interference by RF, light, or particles other than charged particles with energies > 1 MeV but require an etching process that makes time resolved particle detection difficult. Therefore, we have chosen a second diagnostic, the Geiger Muller (GM) detector with mica windows (LND model no. 712) which detect energetic particles in the time scales of microsecond and nanosecond [16]. The principle of this detector is its sensitivity to ionizing radiation and strong discrimination against visible light and other electromagnetic sources that cannot produce ionization on account of its strong electromagnetic shielding. Each detection event, a strong pulse signal of about 5.8 V, 50 μs, is time-stamped registered as a waveform recorded by a PicoScope 5444D digital oscilloscope. Furthermore, the GM signals are calibrated by alphas signals coming from $^{241}$Am and $^{210}$Po in pulse height and duration. GM data provides the number of counts detected in one second time intervals during a run as shown in Fig. 4. In another example experiment, a 10 s run recorded 48 counts per minute (CPM), above the average 12.8 background CPM.

Due to the quiescent plasma conditions with a dominant population of negative ions, MeV alphas from fusion can be distinguished from background signals by using time correlation techniques. Only the GM counts that are synchronized with a strong waveform signal will be selected and retained. This ensures a high signal to noise ratio for the fusion output.

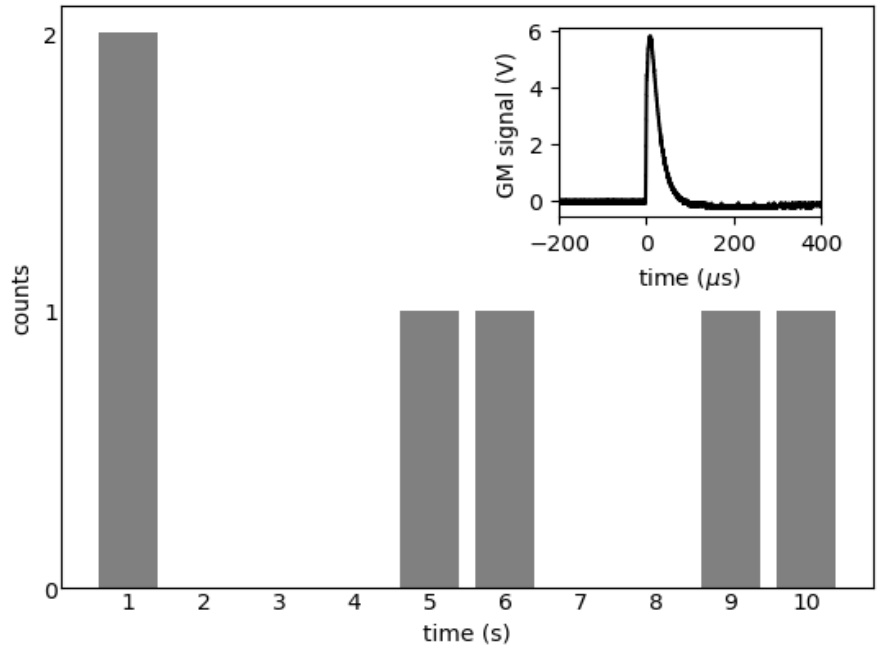

FIG. 4. Counts of individual pulses from GM alpha detector, with 1 second time resolution, during a single 10 second, 19 A plasma discharge (Run 20200819 Shot 7). Inset shows the signal waveform from the GM detector for a typical detection event.

In our experiments, we have found that the input power range is 3 to 4 kW to maximize the number of energetic particles at a pressure of 1 to 1.4 Torr. This is due to the fact that the maximum production of H$^-$ occurs at plasma temperatures below 0.7 eV as found in our previous experiments [2,3]. Figure 5(a) shows the counts observed on a GM detector during a 5-second pulse. In conjunction, spectral data are measured from a Thorlabs compact spectrometer (CCS100). In Fig. 5(b), the total integrated intensity (sum of intensity at all wavelengths measured by spectrometer) is plotted as a function of time, showing that most of the GM counts that are observed during the first 2 to 3 seconds of the pulse correspond to a period of increasing electron emission and plasma formation.

During this same time period, Fig. 5(c) shows that the intensity ratio of the hydrogen Balmer alpha:beta peaks is approximately ~6.8 ± 1.0 in an uncooled system. Assuming local thermodynamic equilibrium, this corresponds to an electron temperature of ~0.58 ± 0.08 eV, allowing for the formation of H$^-$. An additional example of the correlation between GM counts and the plasma temperature can be found in the Supplemental Information [17].

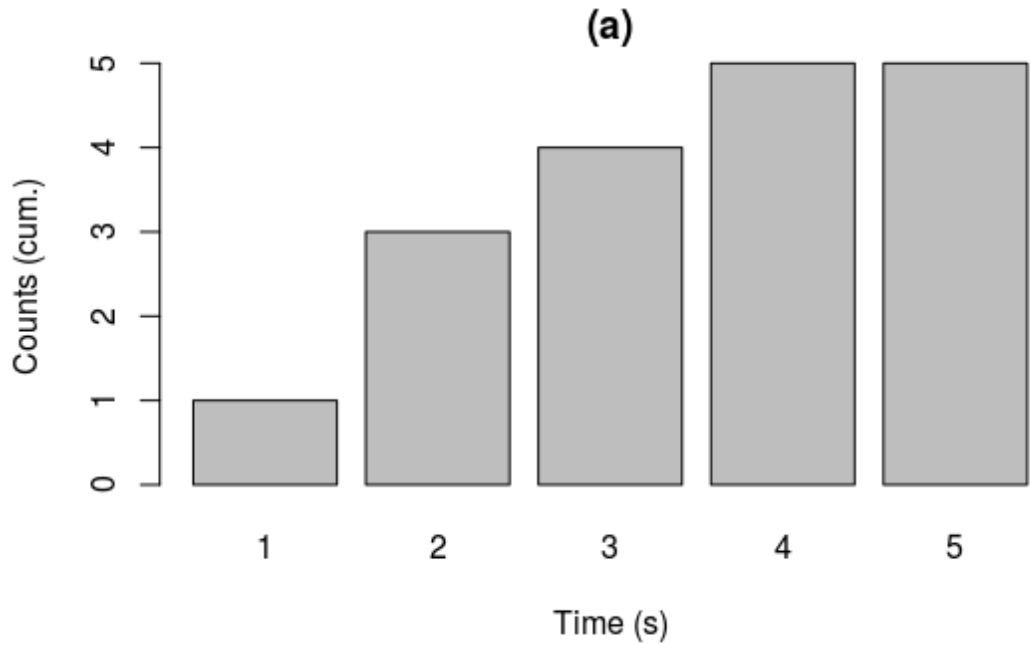

FIG. 5. (a) Cumulative counts observed on GM detector during a 19A current pulse (Run 20200803 Shot 7).

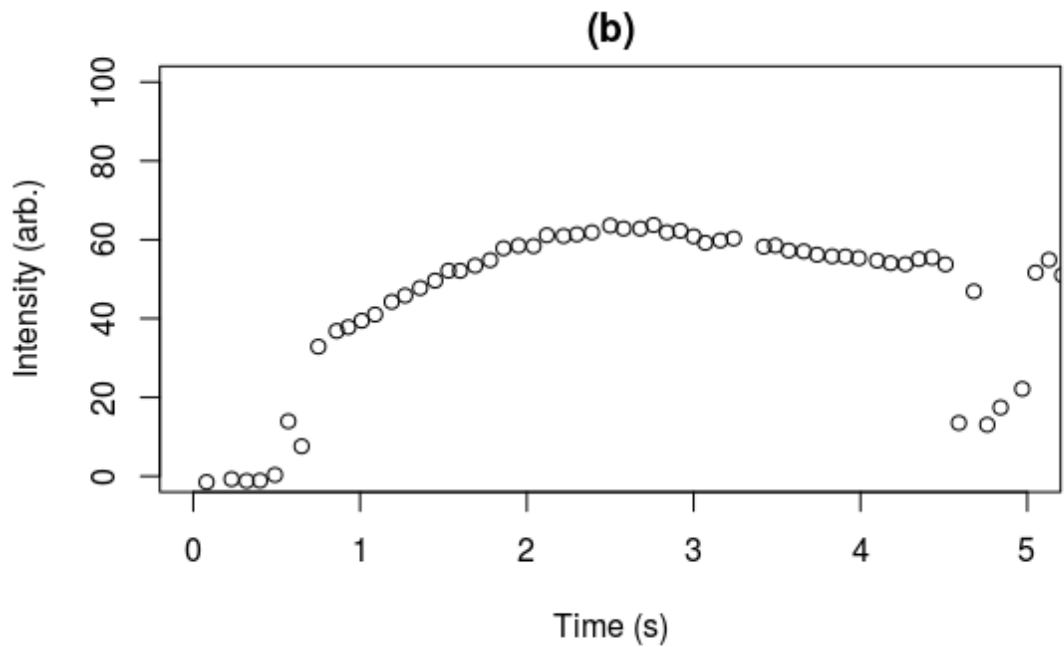

FIG. 5. (b) Plasma formation as indicated by increasing spectral intensity during a pulse, measured by a Thorlabs spectrometer (CCS100).

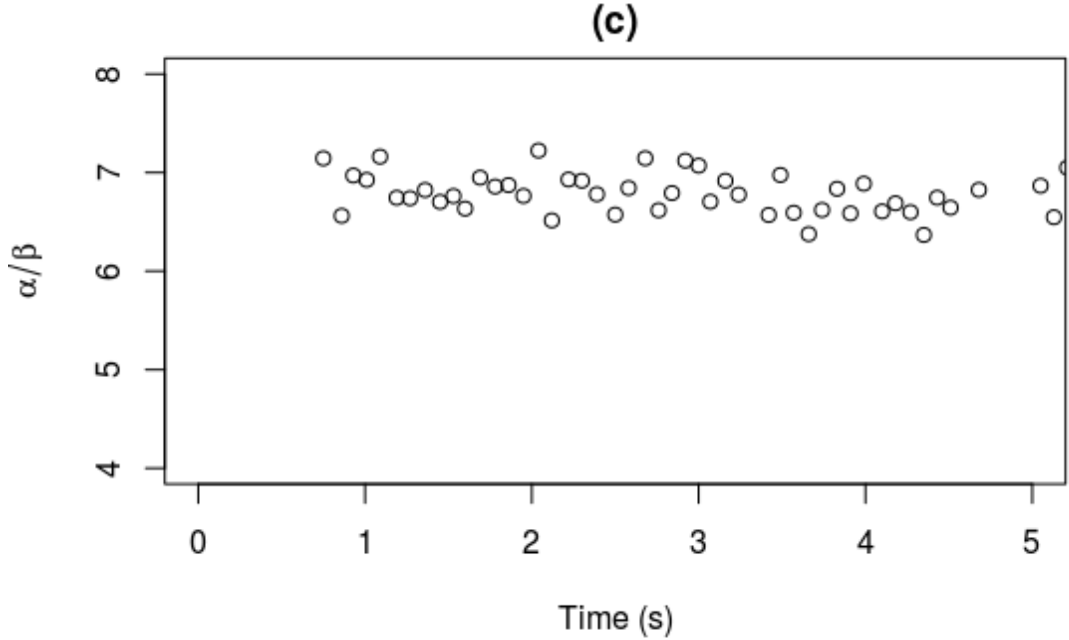

FIG. 5. (c) The ratio of Balmer alpha line 656 nm intensity to beta line 486 nm intensity during the first 5s of a pulse.

While the foregoing experiments with short pulses of 10 s were conducted, we have also conducted experiments in the steady-state regime (~75 s after initiation) with an input power dropping to ~15% of the initial power. Because of this stable plasma regime, MeV alphas from fusion can be distinguished from background signals by using a pulse correlation method. In an example experiment, when the current was pulsed to 23A from the base 16A for 20s, the counts per minute recorded on the GM detector doubled from 14.4 to 31.4 CPM as shown in Fig.6. This is the first time when a fusion plasma is operated in a manner when the regime of each parameter such as temperature, density, RF and DC excitations can be explored.

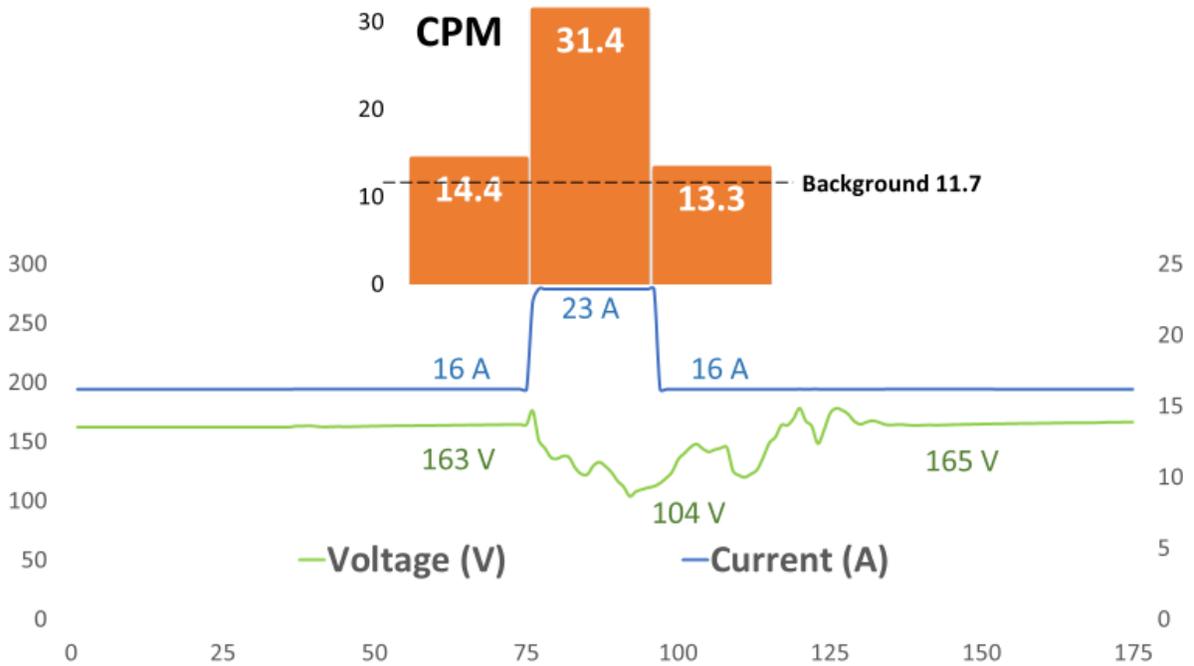

FIG. 6. Correlation of MeV counts and current during steady-state operation. The average counts per minute were recorded before, during, and after a 23 A pulse. Error bars are calculated as [sqrt(N)/N] *CPM, where N is the total counts (Run 20200925).

In another AR system, the observed fusion products were further confirmed as alpha particles through measurements by a third-party, Evans Analytical Group [18], who measured the amount of residual helium in the outer electrode with an operation time of ~3000 minutes. The presence of helium is attributed to accumulated fusion products that are readily turned into helium when they capture two electrons after being embedded in the electrode. Samples of the tantalum outer electrode, ~1 g each, were analyzed and found to contain over four times the helium from an unused tantalum control sample. See Supplemental Information [17] for additional details about the He analysis process.

## CONCLUSION

We have shown in a low temperature plasma the principle that the Coulomb Barrier can be lowered by negative charges. The quiet, weakly-ionized low-temperature plasma can indeed be used to optimize various parameters to yield a high fusion gain. This AR fusion device also illustrates that stability can be achieved with the use of high density of neutrals. This is in contrast to a fully ionized plasma where high-density space charges lead to instabilities.

One theoretical explanation of our results is that the negative ion H⁻ oscillates with the positive boron ions (B⁺). The spatial and temporal variation of the negative ion density can be written as:

$$n(x,t) = n_0 + n_1 \cos(kx)\cos(\omega t) \qquad (3)$$

Here $k = 2\pi/\Lambda$ is the oscillation wave number, $\omega = n_0 e^2/m\varepsilon_0$ is the resonant frequency, $n_1$ is the amplitude of the density fluctuation and $n_0$ the average ion density. The density variation induces oscillating electric field E and the associated potential U as:

$$E(x,t) = \frac{en_1}{k\varepsilon_0}\sin(kx)\cos(\omega t) = E_0 \sin(kx)\cos(\omega t) \qquad (4)$$

$$U(x,t) = \frac{e^2 n_1}{k^2 \varepsilon_0}\cos(kx)\cos(\omega t) = U_0 \cos(kx)\cos(\omega t) \qquad (5)$$

The effective screening potential $U_s$ acts on the group of ions participating in the oscillation and is reduced by the factor of $n_1/n_0$ from $U_0$. The effective barrier reduction of the Coulomb potential can therefore be written as:

$$U_s = \frac{n_1}{n_0}U_0 = \frac{e^2 \Lambda^2 n_1^2}{4\pi^2 \varepsilon_0 n_0} \approx 4.6 \times 10^{-10}[\text{eV} \cdot \text{m}]\frac{\Lambda^2 n_1^2}{n_0} \qquad (6)$$

assuming the density fluctuation is 10% ($n_0 = 10^{23}$ m⁻³ and $n_1 = 10^{22}$ m⁻³) and the oscillation wavelength is 200 μm. The peak screening energy during the oscillation can be calculated to be $U_s \approx 18.4$ keV. The corresponding p-B reaction cross section as calculated by the Schrodinger equation is approximately $10^{-39}$ m$^2$, as shown in Fig. 7. This screening energy is high enough to enhance the p-B fusion reaction rate.

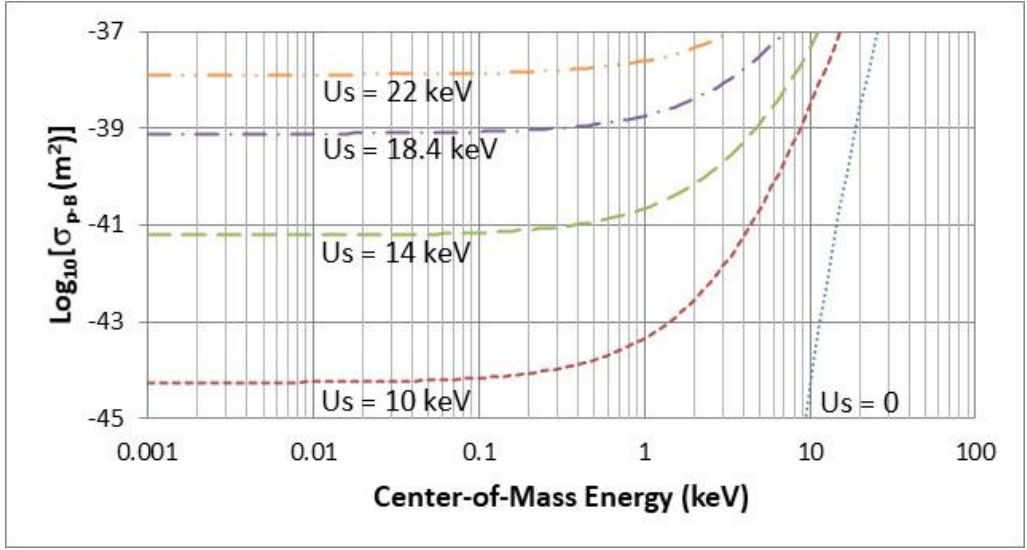

FIG. 7. Cross sections of the p-B fusion reaction with given screening energies: $U_s = 0$, 10, 14, 18.4, and 22 keV. Note the cross section is near $10^{-39}$ m$^2$ at $U_s = 18.4$ keV.

By extrapolating the number of alphas recorded by the GM detectors, the fusion cross section is estimated to be $10^{-39}$ m$^2$. This agrees with the theoretical estimate based on resonant electric fields generated by the oscillations between hydrogen anions formed in the plasma and boron nuclei in solid LaB$_6$, with estimated densities of $\sim 10^{22}$/m$^3$ for H$^-$ and B$^+$ [4].

The present experiments allow the understanding of how negative hydrogen ions might lead to fusion on the solar surface which houses a 500 Km thick layer of negative hydrogen ions with density of $10^{18}$ cm$^{-3}$ at a temperature below 6000 K.

These methods and results presented here portend a new approach to achieve practical nuclear fusion now on our planet earth. Towards that objective, we are optimizing our systems to enhance these collective ion oscillations. Table-top experimental devices presently enable a rigorous scientific foundation for engineering reactors at multiple scales.

## SUPPLEMENT MATERIALS

**MeV Particle Counts and Plasma Temperature**

An example of the correlation between the counts observed on the GM detectors is shown in the experiment in Fig. S1 (a), and the plasma temperature as reported by the Balmer alpha:beta ratio in Fig. S1 (b). The alpha:beta ratio increases to approximately 8 at ~ 6 seconds and 8 seconds into the discharge, corresponding to increased counts observed on the GM detector during the same time period. The increase in alpha:beta ratio indicates a cooling of the plasma, which will favor negative ion production.

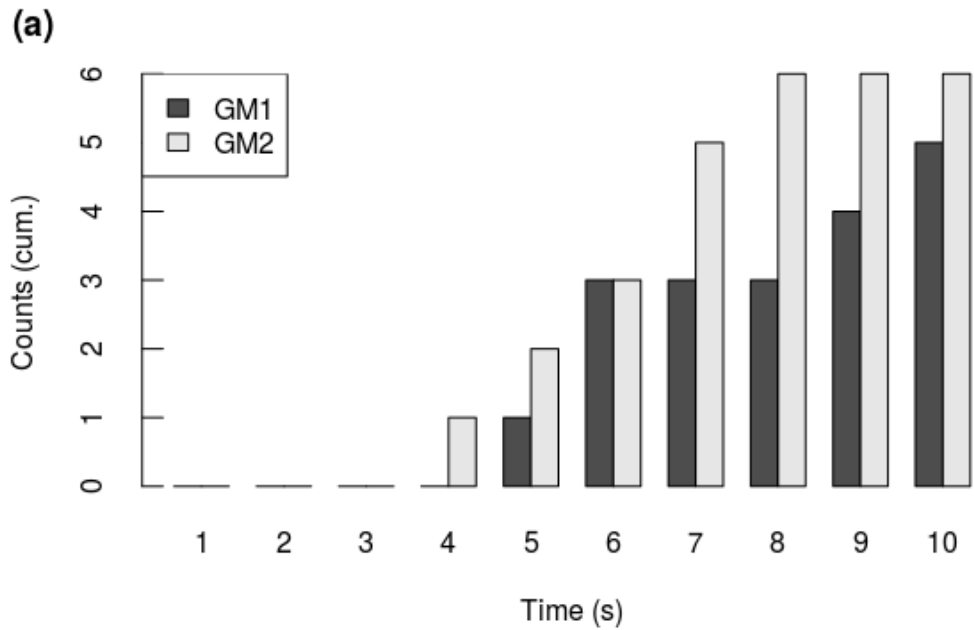

FIG. S1. (a) GM counts observed over 10 second pulse on two separate detectors.

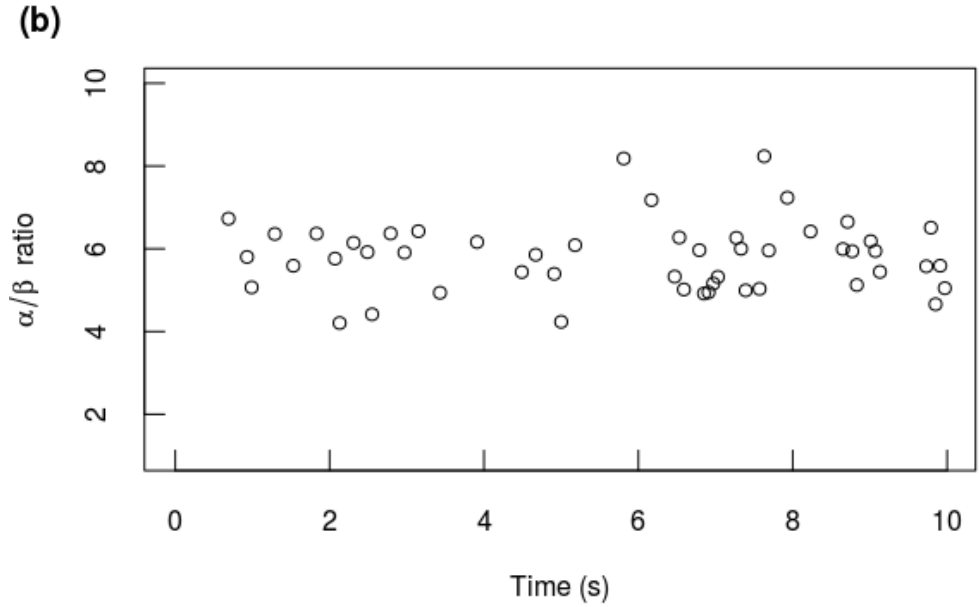

FIG. S1. (b) Ratio of Balmer alpha:beta line intensity during measurement.

**Helium Gas Collection Procedure**

Two samples were analyzed by Evans Analytics Group (EAG). Both were from the same tantalum stock material used for the outer electrode in the system described. The sample labeled "used" in Fig. S2 (a) was cut from the outer electrode of a system that was operated over the course of several months, ½ hour or more at a time, to generate larger quantities of fusion products. The sample labeled as "new" was not placed in the plasma chamber. There was no other source of helium gas introduced, so the significant increase in helium measured is attributed to the accumulation of alpha particles in the electrode material. Figure S2 (b) shows the quantity of helium detected from the used and a new sample of tantalum shroud material. A thermal desorption analysis method to capture helium outgassed from the heated samples.

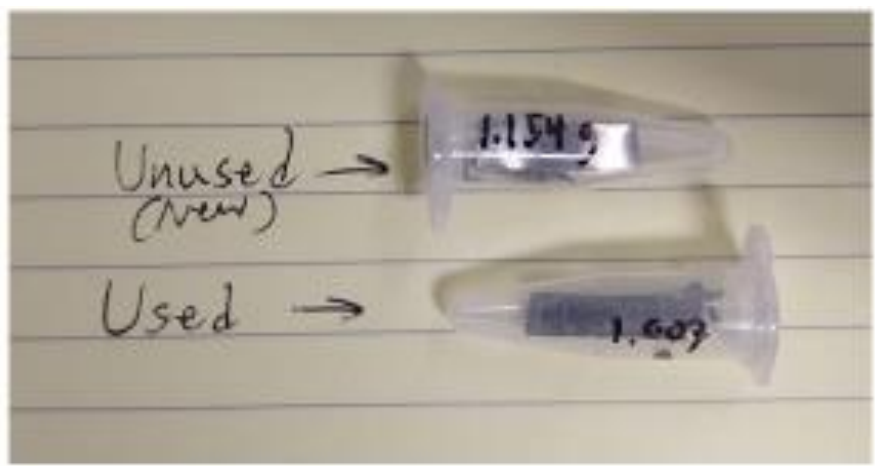

FIG. S2. (a) Tantalum shroud samples for residual Helium gas analysis by EAG [18]. Samples were taken from a reactor with hours of cumulative experiments.

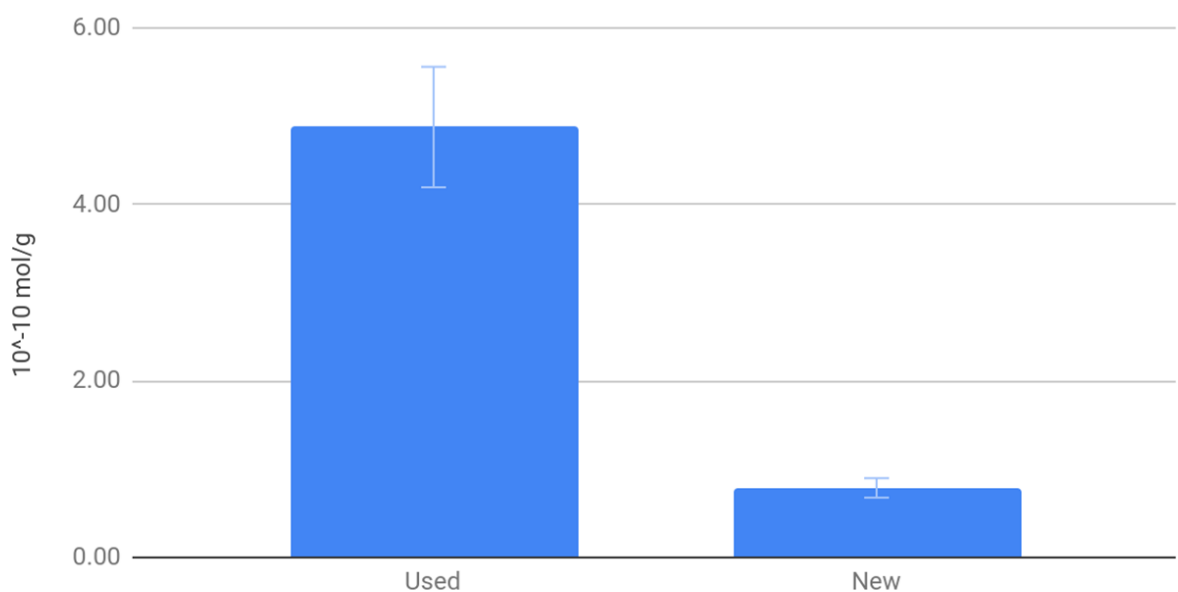

FIG. S2. (b) Quantity of helium detected from a used and a new sample of tantalum shroud by Evans Analytical Group [18] using a thermal desorption analysis method. The reported values in moles of helium measured per gram of sample were $4.88\times10^{-10}$ mol He/g for the experimental sample (used) and $7.87\times10^{-11}$ mol He/g for the control sample (new) with a relative uncertainty of 15%.

# ACKNOWLEDGMENT

We wish to thank CNSI NanoPico facility, UCLA campus for the use of their analytical equipment and their advice, and Drs. Allan Chen and Ted Cremer for discussions and efforts in the development of our experiments and theories. This research is supported by the Independent Research Fund of Alpha Ring International, Ltd. The encouragement given by Peter Liu and Shahi Ghanem is gratefully acknowledged.